\documentclass[article,onecolumn,groupedaddress,showpacs,showkeys,superscriptaddress,10pt]{revtex4-1}
\usepackage[utf8]{inputenc}
\usepackage{graphicx}
\usepackage[colorlinks]{hyperref}
\usepackage{amssymb}
\usepackage{ctable}
\usepackage{amsmath}
\usepackage{amsfonts}
\usepackage{amssymb}
\usepackage{subfigure}
\begin{document}

\title{Squeezing lights via a levitated cavity optomechanics}

\author{Guoyao Li}
	\affiliation{Center for Quantum Technology Research and Key Laboratory of Advanced Optoelectronic Quantum Architecture and Measurements (MOE), School of Physics, Beijing Institute of Technology, Beijing 100081, China}

\author{Zhang-qi Yin}\email{zqyin@bit.edu.cn}
	\affiliation{Center for Quantum Technology Research and Key Laboratory of Advanced Optoelectronic Quantum Architecture and Measurements (MOE), School of Physics, Beijing Institute of Technology, Beijing 100081, China}
	\affiliation{Beijing Academy of Quantum Information Sciences, Beijing 100193, China}
		
\date{\today}

\begin{abstract}
Squeezing light is a critical resource in both fundamental physics and precision measurement. The squeezing light has been generated through optical-parametric amplification inside an optical resonator. 
However, preparing the squeezing light in an optomechanical system is still a challenge for the thermal noise inevitably coupling to the system. 
We consider an optically levitated nano-particle in a bichromatic cavity, in which two cavity modes could be excited by the scattering photons of the dual-tweezers respectively.
Based on the coherent scattering mechanism, the ultra-strong coupling between the cavity field and torsional motion of nano-particle could be achieved for the current experimental conditions.
With the back-action of the optically levtiated nano-particle,
the broad single-mode squeezing light can be realized in the bad cavity regime.
Even at room temperature, the single-mode light can be squeezed for more than 17 dB, which is far beyond the 3 dB limit.
The two-mode squeezing lights can also be generated, if the optical tweezers contain two frequencies, one is on the red sideband of the cavity mode, the other is on the blue sideband. 
The two-mode squeezing can be maximized near the boundary of the system stable regime, and is sensitive to both the cavity decay rate and the power of the optical tweezers.
\end{abstract}

\maketitle

\section{Introduction}

Squeezing light, in which the quantum fluctuation is modulated below the shot noise level, has been regarded as a powerful resource in fundamental physics, e,g. improving the sensitivity of gravitational wave detection~\cite{Aasi_2013}, cooling the motion of a macroscopic mechanical object below the quantum backaction limit~\cite{clark2017sideband},  engineering matter interactions~\cite{zeytinouglu2017engineering}, and inducing the topological phase transitions~\cite{peano2016topological}, among many others.
The squeezing light can be generated via the nonlinear optics, such as parametric down-conversion process~\cite{1994PhRvA..49.1337F,1994PhRvA..49.4055M,wu1986generation,aspelmeyer2014cavity}. 
The motion of the mechanical oscillator in cavity optomechanical systems could be regarded as an effective nonlinear optical medium, generating the pondermotive squeezing state by the back-action interaction~\cite{PhysRevX.3.031012,ockeloen2017noiseless,aspelmeyer2014cavity,pontin2014frequency}.
With various optomechanical coupling mechanism, cavity optomechanical system produces abundant squeezing light sources~\cite{PhysRevA.79.024301,sainadh2015effects,pontin2014frequency,zippilli2015entanglement,kronwald2014dissipative,pontin2014frequency}, even in pulse driving regime~\cite{pontin2014frequency,kronwald2014dissipative}.
However, due to the noise inevitably induced from the thermal bath, as well as the limitation of the nonlinear interaction, generating the substantial pondermotive squeezing light is still a challenge for the optomechanical system~\cite{PhysRevX.3.031012}. 

The levitated optomechanical system has raised widespread interest in macroscopic quantum superposition~\cite{yin2013large,romero2011large,Chen2018}, quantum time crystal~\cite{PhysRevA.102.023113,Huang2018}, and quantum information processing~\cite{zhang2021quantum}. 
The squeezing has been studied in the levitated optomechanical system, in which the nano-particle is optically levitated and coupled with a cavity field~\cite{chang2010cavity}.
Under ultra-high vacuum (gas pressure $p=10^{-10}$ Torr),
the squeezing light is estimated over $15$ dB below the noise level~\cite{chang2010cavity}.
As for the experiment, the motional state for the optically levitated nano-particle has been squeezed up to $2.7$ dB~\cite{rashid2016experimental}.
Recently, the coherent scattering mechanism has been proposed \cite{gonzalez2019theory}, and experimentally realized to cool the nano-particle to quantum ground state ~\cite{delic2020cooling,magrini2021real,Tebbenjohanns2021quantum,schafer2021cooling}.
Moreover, the strong coherent coupling between the cavity field and the motion of the nano-particle has also been observed in experiment~\cite{de2021strong}.
With the attainable ultra-high quality factor (beyond $10^9$) of the optically nano-particle \cite{asenbaum2013cavity,millen2020optomechanics,jain2016direct}, 
it is expected that the stronger and robust squeezing light can be realized via the coherent scattering~\cite{gonzalez2019theory,PhysRevResearch.2.013052,millen2020optomechanics,yin2013optomechanics,chang2010cavity}.

In this paper, we propose a scheme to generate the single/two-mode squeezing light source via a vacuum levitated optomechanical system, in which the nano-particle is optically levitated by the dual-tweezers~\cite{gonzalez2019theory,chang2010cavity}.
Based on the coherent scattering mechanism~\cite{gonzalez2019theory}, we find that the ultra-strong coupling between the cavity mode and the torsion mode of the nano-particle is available by the current experimental parameters~\cite{kockum2019ultrastrong,li2021steady,hoang2016torsional,ahn2018optically}.
In order to obtain the squeezed single-mode light in steady state, the optical tweezers is detuned to the red sideband of the cavity mode.
By the back-action of the optically levitated nano-particle~\cite{pontin2014frequency}, the single-mode light can be squeezed over 17 dB.
The strong and broad single-mode squeezing light is observed in a bad cavity.
Furthermore, in order to generate the two-mode squeezing light, we consider the optical tweezers with two frequencies, one is on the red sideband of the cavity mode, the other is on the blue sideband.
We find that the two-mode squeezing can be maximized when the dynamics closes to the system instability~\cite{genes2008robust}.
The squeezing is sensitive to both the cavity decay rate and optical tweezers power.

The paper is organized as follows. The model and the system dynamics are depicted in Section~\ref{II}. Then the results of one-mode squeezing light and two-mode squeezing light are discussed in Sections~\ref{III} and~\ref{IV}. At last, a brief conclusion is given in Section~\ref{V}.
\section {Model \& Dynamics} \label{II}

\begin{figure}
  \centering
  \includegraphics[width=12cm]{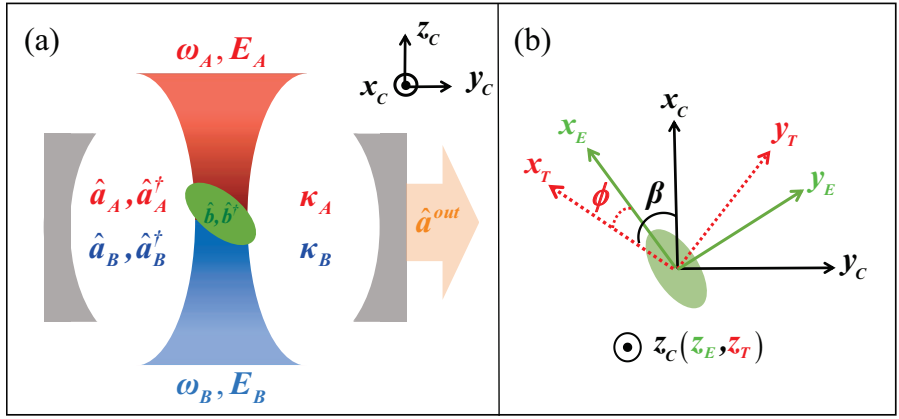}

  \caption{(Color online) (a) Schematic diagram of the levitated optomechanical systems. The nano-ellipsoid is placed into a bichromatic cavity and optically levitated by the dual-tweezers with two frequencies $\omega_A$ and $\omega_B$, and amplitudes $E_A$ and $E_B$, respectively. Two cavity modes $\hat a^\dag_A$ ($\hat a_A$) and $\hat a^\dag_B$ ($\hat a_B$) are excited by the scattering photons with the decay rates $\kappa_A$ and $\kappa_B$. (b) The orientation of the nano-ellipsoid $\{x_E,y_E,z_E\}$ rotates under the tweezers coordinate  $\{x_T,y_T,z_T\}$ with a small angel $\phi$. $\beta$ is the angel between the cavity coordinate axis $x_C$ and tweezers coordinate axis $x_T$. Tweezers propagate along the direction of axis $z$.}
  \label{Fig:1}
\end{figure}

We consider a uniform isotropic non-dispersive nano-ellipsoid, which is optically levitated by linearly polarized dual-tweezers (A and B) in a bichromatic cavity as shown in Figure~\ref{Fig:1}a.
The optical tweezers A (B) with frequency $\omega_{A(B)}$ is on the red (blue) sideband of cavity mode $\hat a_A$ ($\hat a_B$).
Under this condition, the nano-ellipsoid will be cooled and heated by the optical tweezers A and B, respectively.
If the cooling rate is larger than the heating rate, the motional state can be cooled down to a low temperature and the system remains dynamically stable.

The motion of the levitated nano-ellipsoid is characterized by the five degrees of freedom where $\{x,y,x\}$ for the center-of-mass motion in position $\hat R$ and $\{\theta,\phi\}$ for the torsional motion in orientation $\hat \Omega$, as shown in Figure~\ref{Fig:1}b.
By locating the nano-ellipsoid to the node (anti-node) of the cavity modes, the center-of-mass motion and torsional motion can be decoupled from each other~\cite{li2021steady}.
Besides, intrinsic optomechanical coupling between the cavity mode and torsional motion of the nano-ellipsoid is typically much weaker than the coherent scattering coupling strength~\cite{li2021steady,delic2020cooling,gonzalez2019theory}.
Therefore, we only consider the coherent scattering coupling between the cavity mode and the motion of the nano-ellipsoid.
Without loss of generality, we take the torsional motion as an example in the following discussion~\cite{hoang2016torsional,ahn2018optically}.

\begin{figure}
  \centering
  \includegraphics[width=12cm]{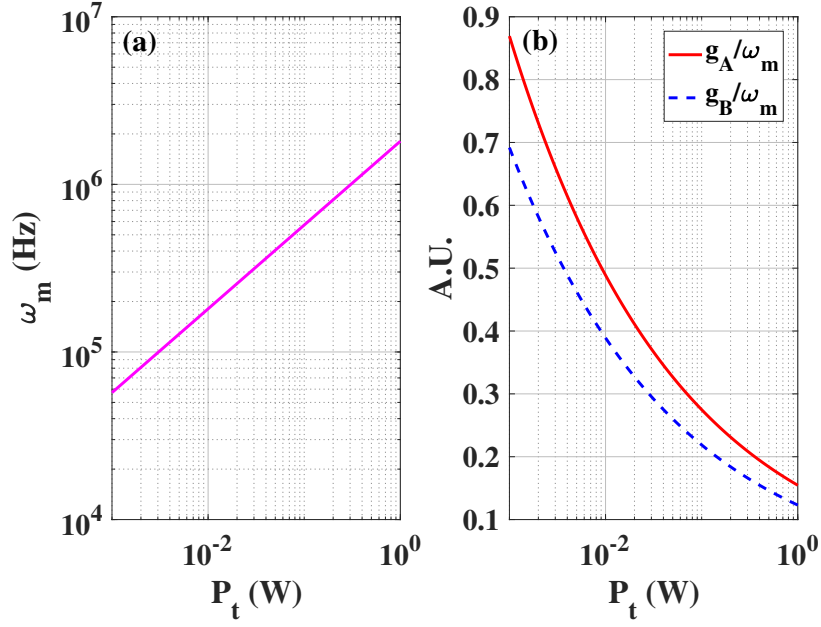}

  \caption{(Color online) The torsional frequency $\omega_m$ (a) and the ratio between the coherent scattering coupling and torsional frequency (b) as a function of the tweezers power $P_t$. The parameters are given as follows: the cavity length $L=1$ mm, the wavelength of the optical tweezers $\lambda_A=780$ nm ($\lambda_B=980$ nm), the beam waist of the tweezers in focus $w^j_0=1$~$\mu$m, the principle axes of the nano-ellipsoid $a=2b=2c=100$ nm, the relative permittivity of the nano-ellipsoid $\varepsilon=2.1$, the density of the nano-ellipsoid $\rho=2200$ kg/m$^3$. For convenience, the power of two optical tweezers are assumed the same $P_j=P_t$.}
  \label{Fig:2}
\end{figure}

With respect to the interaction picture $H_0= \hbar\sum\limits_{j = A,B}{{\omega _j}} a_j^\dag {a_j}$, the interaction Hamiltonian for the system can be written as~\cite{gonzalez2019theory}
\begin{equation}
\hat H{\rm{ = }}\hbar \sum\limits_{j = A,B} \Delta _c^j\hat a_j^\dag {\hat a_j} + \hbar {\omega _m}{\hat b^\dag }\hat b - \hbar \sum\limits_{j = A,B} {{g_j}\left( {\hat a_j^\dag  + {{\hat a}_j}} \right)\left( {{{\hat b}^\dag } + \hat b} \right)},
\label{EQ1}
\end{equation}
where $\hat a^\dag_j$ and $\hat a_j$ ($\hat b^\dag$ and $\hat b$) are the bosonic creation and annihilation operators with the commutation relations $\left[ {\hat a_j,{\hat a^{\dag}_j}} \right] = 1$ $(\left[ {{\hat b },\hat b^{\dag}} \right] = 1)$~\cite{vitali2007optomechanical},
$\Delta^j_c=\omega^j_c-\omega_j$ is the detuning between the cavity frequency $\omega^j_c$ and the optical tweeezers frequency $\omega_j$. 
$\omega_m$ is the torsional mode frequency,  and $g_j$ is the coherent scattering coupling. They can be calculated by (see Appendix~\ref{AP_B})
\begin{equation}
 {g_j} = \left( {{\alpha _a} - {\alpha _b}} \right){E^j_0}{\xi _0}\cos \left( \varphi  \right)\sqrt {\frac{{{\omega^j _c}}}{{8\hbar {\varepsilon _0}{V^j_c}}}}
\label{EQ2}
\end{equation}
and
\begin{equation}
 {\omega _m} = \sqrt {{{ \sum\limits_{j = A,B} E_0^{j2} \left( {{\alpha _a} - {\alpha _b}} \right)} \mathord{\left/
 {\vphantom {{\left( {E_0^AE_0^A + E_0^BE_0^B} \right)\left( {{\alpha _a} - {\alpha _b}} \right)} {2I}}} \right.
 \kern-\nulldelimiterspace} {2I}}}
 \label{EQ3}
\end{equation}
respectively.
For the current experimental parameters, the torsional frequency can be in the order of MHz as shown in Figure~\ref{Fig:2}a. 
Interestingly, the ratio ($g_j/\omega_m$) between the coherent scattering coupling and the torsional frequency is larger than $0.1$ as shown in Figure~\ref{Fig:2}b, in which the ultra-strong optomechanical coupling regime has been reached ~\cite{kockum2019ultrastrong}.
In this regime, both of the rotating and anti-rotating wave terms have to be considered.
As $\hat a_j$ ($\hat a_j^\dag$) depends on $\hat b+\hat b^\dag$, one field quadrature of the cavity mode will be squeezed, while another will be anti-squeezed~\cite{1994PhRvA..49.1337F,1994PhRvA..49.4055M}.

Derived from the system Hamiltonian~\eqref{EQ1}, the dynamics of the system is characterized by the following quantum Langevin equations:
\begin{equation}
\frac{{d\hat b}}{{dt}} =  - \left( {\frac{{{\gamma _m}}}{2} + i{\omega _m}} \right)\hat b + i\sum\limits_{j = A,B} {g_j}\left( {\hat a^\dag_j  + \hat a_j} \right) + \sqrt {{\gamma _m}} {\hat b^{in}},
\label{EQ4}
\end{equation}
\begin{equation}
\frac{{d{{\hat a}_j}}}{{dt}} =  - \left( {\frac{{{\kappa _j}}}{2} + i{\Delta _j}} \right){\hat a_j} + i{g_j}\left( {{{\hat b}^\dag } + \hat b} \right) + \sqrt {{\kappa _j}} {\hat a^{in}},
\label{EQ5}
\end{equation}
where $\gamma_m$ is the damping rate of the torsional mode.
Noted that the operators are already replaced by $\delta \hat O\rightarrow\hat O$, where $\delta \hat O$ denotes the fluctuation of $\hat O=\{\hat a_j,\hat b$\}.
$\hat a ^{in}$ and $\hat b ^{in}$ are the zero-mean Gaussian noises, satisfying the correlation relations
$
\left\langle {{\hat{a}^{in}}\left( t \right){\hat{a}^{in\dag }}\left( {t'} \right)} \right\rangle  = \delta \left( {t - t'} \right)
$
and
$
\left\langle {{\hat{b}^{in}}\left( t \right){\hat{b}^{in\dag }}\left( {t'} \right)} \right\rangle  = \left( {\bar n + 1} \right)\delta \left( {t - t'} \right)
$
respectively~\cite{vitali2007optomechanical}.
Here $\bar n=\{\exp(\hbar \omega_m /k_B T)-1\}^{-1}$ is the mean thermal excitation numbers at bath temperature $T$, and $k_B$ is Boltzmann constant.
The linearized Langevin Equations~\eqref{EQ4} and \eqref{EQ5}  can be
rewritten in the matrix form 
\begin{equation}
\dot{\hat f}(t)=A \hat f(t)+\hat D(t)
\label{EQ6}
\end{equation}
where
$\hat f^T(t)=\{\hat b(t), \hat b^\dag(t),\hat a_A(t), \hat a_A^\dag(t),\hat a_B(t), \hat a_B^\dag(t)\}$ 
and
$\hat D^T(t)=\{\sqrt{\gamma_m}\hat b^{in}(t),\sqrt{\gamma_m}\hat b^{in\dag}(t),
\\
\sqrt{\kappa_A}\hat a_A^{in}(t),\sqrt{\kappa_A}
\hat a_A^{in\dag}(t),
\sqrt{\kappa_B}\hat a_B^{in}(t),\sqrt{\kappa_B}\hat a_B^{in\dag}(t)\}$.
The drift matrix $A$ is given by

\begin{equation}
A=%
\begin{pmatrix}
\begin{array}{cccccc}
-\frac{\gamma}{2}-i\omega_m  & 0  & ig_A & ig_A & ig_B & ig_B \\
0  & -\frac{\gamma}{2}+i\omega_m  & -ig_A & -ig_A & -ig_B & -ig_B  \\
ig_A & ig_A & -\frac{\kappa_A}{2}-i\Delta_A & 0 & 0 & 0\\
-ig_A & -ig_A & 0 & -\frac{\kappa_A}{2}+i\Delta_A & 0 & 0\\
ig_B & ig_B & 0 & 0 & -\frac{\kappa_B}{2}-i\Delta_B & 0\\
-ig_B & -ig_B & 0 & 0 & 0 & -\frac{\kappa_B}{2}+i\Delta_B\\
\label{EQ7}
\end{array}%
\end{pmatrix}%
.
\end{equation}

After the Fourier transform
$\hat O\left( t  \right) = \int_{ - \infty }^{ + \infty } {\hat O \left( \omega \right)} {e^{ - i\omega t}}d\omega$,
the steady state solutions of the Langevin Equations~\eqref{EQ4} and~\eqref{EQ5} in the frequency domain are 
\begin{equation}
\hat b\left( \omega  \right) = \frac{{i{g_j}\left( {\hat a_j^\dag \left( { - \omega } \right) + \hat a_j\left( \omega  \right)} \right) + \sqrt {{\gamma _m}} {{\hat b}^{in}}}}{{\frac{{{\gamma _m}}}{2} + i\left( {{\omega _m} - \omega } \right)}},
\label{EQ8}
\end{equation}
\begin{equation}
{\hat a_j}\left( \omega  \right) = \frac{{i{g_j}\left( {{{\hat b}^\dag }\left( { - \omega } \right) + \hat b\left( \omega  \right)} \right) + \sqrt {{\kappa _j}} \hat a_j^{in}}}{{\frac{{{\kappa _j}}}{2} + i\left( {{\Delta _j} - \omega } \right)}},
\label{EQ9}
\end{equation}
where the corresponding correlation noises are 
$
\left\langle {{{\hat b}^{in}}\left( \omega  \right){{\hat b}^{in\dag }}\left( { - \omega '} \right)} \right\rangle  = 2\pi \left( {2\bar n + 1} \right)\delta \left( {\omega  + \omega '} \right)
$
and
$
\left\langle {\hat a_j^{in}\left( \omega  \right)\hat a_{j'}^{in\dag }\left( { - \omega '} \right)} \right\rangle  = 2\pi {\delta _{jj'}}\left( {\omega  + \omega '} \right)
$
, respectively.
Furthermore, Equations~\eqref{EQ8} and \eqref{EQ9} can be rewritten as $\hat A(\omega)=J^{-1}(\omega)\hat B(\omega)$
where
$
\hat A(\omega)=\{\hat b(\omega),\hat b^\dag(-\omega),\hat a_A(\omega),\hat a^\dag_A(-\omega),\\
\hat a_B(\omega),\hat a^\dag _B(-\omega)\}^T
$
,
$
B(\omega)=\{\sqrt{\gamma_m} \hat b^{in}(\omega),\sqrt{\gamma_m} \hat b^{in\dag}(-\omega),\sqrt{\kappa_A} \hat a^{in}_A(\omega),\sqrt{\kappa_A}  \hat a^{in\dag}_A(-\omega),\\ \sqrt{\kappa_B} \hat a^{in}_B(\omega), \sqrt{\kappa_B}  \hat a^{in\dag}_B(-\omega)\}^T
$
,
\begin{equation}
J\left( \omega  \right) = \left[ {\begin{array}{*{20}{c}}
{{u_1}}&0&{ - i{g_A}}&{ - i{g_A}}&{ - i{g_B}}&{ - i{g_B}}\\
0&{{u_2}}&{i{g_A}}&{i{g_A}}&{i{g_B}}&{i{g_B}}\\
{ - i{g_A}}&{ - i{g_A}}&{{v_1}}&0&0&0\\
{i{g_A}}&{i{g_A}}&0&{{v_2}}&0&0\\
{ - i{g_B}}&{ - i{g_B}}&0&0&{{w_1}}&0\\
{i{g_B}}&{i{g_B}}&0&0&0&{{w_2}}
\end{array}} \right],
\label{EQ10}
\end{equation}
$u_1={\frac{{{\gamma _m}}}{2} + i\left( {{\omega _m} - \omega } \right)}$,
$u_2={\frac{{{\gamma _m}}}{2} - i\left( {{\omega _m} + \omega } \right)}$,
$v_1={\frac{{{\kappa _A}}}{2} + i\left( {{\Delta _A} - \omega } \right)}$,
$v_2={\frac{{{\kappa _A}}}{2} - i\left( {{\Delta _A} + \omega } \right)}$,
$w_1={\frac{{{\kappa _B}}}{2} + i\left( {{\Delta _B} - \omega } \right)}$,
and
$w_2={\frac{{{\kappa _B}}}{2} + i\left( {{\Delta _B} - \omega } \right)}$.
Based on the standard input-output relation
\begin{equation}
\hat a ^{out}_j (\omega) =\sqrt{\kappa_j} \hat a_j (\omega)-\hat {a} ^{in}_j (\omega),
\label{EQ11}
\end{equation}
the outputs cavity mode $a ^{out}_j (\omega)$ can be attainable.

\section{Single-mode Squeezing} \label{III}

We consider the case that the nano-ellipsoid is levitated by only one optical tweezers A, while the other optical tweezers B is turned off.
Only one cavity mode $\hat a_A$ is excited by the scattering photons of the optical tweezers A.
From Equations~\eqref{EQ8} and \eqref{EQ9}, the cavity mode $\hat a_A(\omega)$ can be solved as
\begin{equation}
\hat a_A(\omega) = m_1^A(\omega){\hat b^{in}}(\omega) + m_2^A(\omega){\hat b^{in\dag }}(-\omega) + m_3^A(\omega)\hat a_A^{in}(\omega) + m_4^A(\omega)\hat a_A^{in\dag }(-\omega)
\label{EQ12}
\end{equation}
where
\begin{equation}
\begin{array}{l}
m_1^A(\omega) = i\sqrt {{\gamma _m}} {g_A}{u_2}{v_2}/z,\\
\\
m_2^A(\omega)= i\sqrt {{\gamma _m}} {g_A}{u_1}{v_2}/z,\\
\\
m_3^A(\omega) = \sqrt {{\kappa _A}} \{{g^2_A}( {u_1}-{u_2})+u_1 u_2 v_2\}/z,\\
\\
m_4^A(\omega) = \sqrt {{\kappa _A}} {g^2_A}( {u_1}-{u_2})/z,\\
\end{array}
\label{EQ13}
\end{equation}
with $z=g^2_A (u_1-u_2) (v_1-v_2)+u_1u_2v_1v_2$.
Furthermore, substituting  Equation~\eqref{EQ12} to Equation~\eqref{EQ11}, one can obtain the stationary squeezing spectrum of the transmitted field:
\begin{equation}
{S_\vartheta}\left( \omega  \right) = \left\langle {\delta X_\vartheta ^{out}\left( \omega  \right)\delta X_\vartheta ^{out}\left( {\omega '} \right)} \right\rangle
\label{EQ14}
\end{equation}
where $\delta X_\vartheta ^{out}\left( \omega  \right) = {e^{ - i\vartheta }}\delta a_A^{out}\left( \omega  \right) + {e^{i\vartheta }}\delta a_A^{out\dag }\left( -\omega  \right)$ with $\vartheta$ being the measurement phase angle in homodyne detection.
Or more specifically, Equation~\eqref{EQ14} can be rewritten as the form of
${S_\vartheta}\left( \omega  \right) = {S_{\hat a{{\hat a}^\dag }}} + {S_{{{\hat a}^\dag }\hat a}} + \left[ {{e^{ - 2i\vartheta }}{S_{\hat a\hat a}} + c.c} \right]$
where

\begin{equation}
\begin{aligned}
{S_{\hat a{{\hat a}^\dag }}}\left( \omega  \right) =& \left\langle {\delta \hat a_A^{out}\left( \omega  \right)\delta {{\hat a}^{out\dag}_A }\left( {-\omega '} \right)} \right\rangle \\
=& {\kappa _A}\left[m_3^A(\omega)-1/\sqrt {\kappa _A}\right]\left[m_3^{A*}(\omega)-1/\sqrt {\kappa _A}\right] \\
&+{\kappa _A} \left[(\bar n+1) m_1^A (\omega) m_1^{A*} (\omega)+ \bar n m_2^A (\omega) m_2^{A*} (\omega)\right],
\end{aligned}
\label{EQ15}
\end{equation}

\begin{equation}
\begin{aligned}
{S_{{{\hat a}^\dag }\hat a}}\left( \omega  \right) &=\left\langle {\delta {{\hat a}^{out\dag}_A }\left( -\omega  \right)\delta \hat a_A^{out}\left( {\omega '} \right)} \right\rangle \\
&={\kappa _A}\left[ {{{\left| {m_4^A\left( { - \omega } \right)} \right|}^2} + \left( {\bar n + 1} \right){{\left| {m_2^A\left( { - \omega } \right)} \right|}^2} + \bar n{{\left| {m_1^A\left( { - \omega } \right)} \right|}^2}} \right],\\
\end{aligned}
\label{EQ16}
\end{equation}
and
\begin{equation}
\begin{aligned}
{S_{\hat a\hat a}}\left( \omega  \right) =& \left\langle {\delta \hat a_A^{out}\left( \omega  \right)\delta \hat a_A^{out}\left( {\omega '} \right)} \right\rangle 
= {\kappa _A}\left[m_3^A( \omega)-1/\sqrt {\kappa _A}\right]m_4^A(- \omega) \\
&+{\kappa _A} \left[ (\bar n+1) m_1^A( \omega)m_2^A(- \omega) +\bar n m_2^A( \omega)m_1^A(- \omega) \right].
\end{aligned}
\label{EQ17}
\end{equation}
It is noted that the maximum squeezing could reach when $dS_\vartheta(\omega)/\vartheta=0$.
Then it is easy to find out that the maximum squeezing of single output cavity mode is~\cite{sainadh2015effects,zhang2018quantum}
\begin{equation}
{S_1}\left( \omega  \right) = {S_{\hat a{{\hat a}^\dag }}}\left( \omega  \right) + {S_{{{\hat a}^\dag }\hat a}}\left( \omega  \right) - 2\left| {{S_{\hat a\hat a}}\left( \omega  \right)} \right|
\label{EQ18}
\end{equation}
as ${e^{2i\vartheta }} =  - {{{S_{\hat a\hat a}}} \mathord{\left/
 {\vphantom {{{S_{\hat a\hat a}}} {\left| {{S_{\hat a\hat a}}} \right|}}} \right.
 \kern-\nulldelimiterspace} {\left| {{S_{\hat a\hat a}}} \right|}}$.
The output cavity mode is squeezed if $S_1(\omega)$ below the shot-noise level ($S_1(\omega)<1$).
\begin{figure}
  \centering
  \includegraphics[width=10cm]{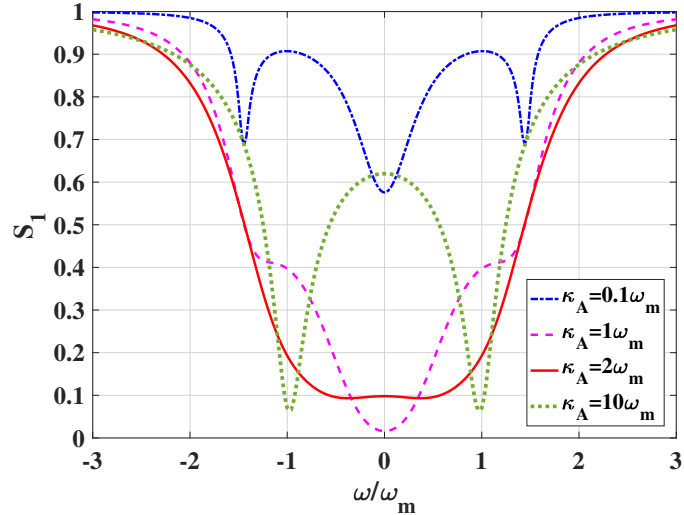}

  \caption{(Color online) $S_1(\omega)$ as a function of the cavity decay rate $\kappa_A$. Parameters are listed as follow: optical tweezers power in focus $P_A=0.05$ W, pressure of the residual gas $p=10^{-4}$ Pa, temperature of the residual gas $T_a=300$ K, bath temperature for the torsional mode $T=300$ K, the accommodation efficient $\gamma_{ac}=0.9$. Other parameters are the same with Figure~\ref{Fig:2}.}
  \label{Fig:3}
\end{figure}

We adopt that the frequency of the optical tweezers A is on the red sideband of the cavity mode, e.g. $\Delta_A=\omega_m$.
In order to maintain the system stability,  the Routh-Hurwitz criterion must be fulfilled~\cite{vitali2007optomechanical}.
As shown in Figure~\ref{Fig:3}, we find that the output cavity mode can be squeezed even under room temperature. 
Obviously, the optimal squeezing of the output cavity field occurs around resonance regime $\omega=0$, where the dissipative part of the mechanical susceptibility has no response in dynamics~\cite{1994PhRvA..49.4055M}.
In the good cavity regime $\kappa_A<\omega_m$, the ponderomotive squeezing will be suppressed ~\cite{kronwald2014dissipative}.
As the cavity decay rate increases from $\kappa_A=0.1\omega_m$ to  $\kappa_A=1\omega_m$, the minimum of $S_1(\omega)$ decreases. 
The output cavity mode can be squeezed up to $17$ dB when the life time of cavity photons approximately equals to a period of torsional motion ($\kappa_A \approx \omega_m$).
As the cavity decay rate continues to increase, the broadband and strong squeezing generates for a bad cavity condition $\kappa_A \approx 2\omega_m$.
Then the broadband squeezing will split into two dips under the bad cavity regime, and the squeezing degree will reduce at $\omega=0$ for the larger dissipation~\cite{kronwald2014dissipative}.

\begin{figure}
  \centering
  \includegraphics[width=10cm]{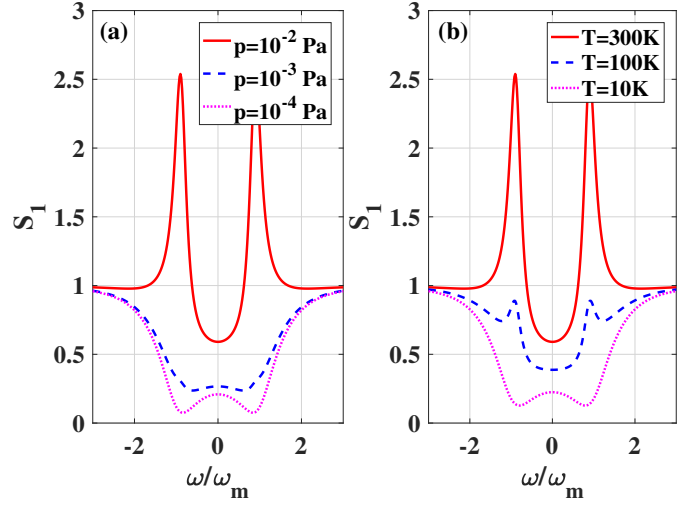}

  \caption{(Color online) $S_1(\omega)$ as a function of the pressure of the surrounding gas $p$ (a) and the temperature of torsional mode $T$ (b). In pictures (a), the temperature of the torsional mode is assumed $T=300$ K, while the pressure of residual gas is set to $10^{-2}$ Pa in picture (b).
  The cavity decay rate is $\kappa_A=3\omega_m$ in both picture (a) and (b).
  Other parameters are the same with the Figure~\ref{Fig:3}.}
  \label{Fig:4}
\end{figure}

Noted that the model we proposed is an open system, in which the optically levitated nano-ellipsoid irreversibly couples to the environment by the collision of the surrounding gas~\cite{jain2016direct}. 
With the back-action interaction, the thermal disturbance could act on cavity mode, leading to the decoherence.
Here we investigate the effects on the squeezing lights from the perspective of the surrounding gas pressure and temperature of the torsional motion.
As shown in Figure~\ref{Fig:4}a, $S_1(\omega)$ depends on the gas pressure $p$.
The lower pressure of the surrounding gas leads to the higher squeezing lights.
However, limited by the stability in ultra-high vacuum and the photons recoiling~\cite{jain2016direct}, the squeezing threshold exists.
On the other hand, the higher temperature for the torsional mode will extremely enlarge the thermal phonons number, which would destroy the squeezing state of the output lights.
Therefore, many schemes based on the optomechanics require the low temperature environment~\cite{PhysRevX.3.031012,pontin2014frequency,rashid2016experimental}.
In order to realize the stronger squeezing light, the high quality factor and low temperature environment is necessary for the levitated optomechanical system~\cite{pontin2014frequency}.
For our scheme, the squeezing is robust to the thermal environment as shown in in Figure~\ref{Fig:4}b.

\section{Two-mode Squeezing} \label{IV}

Our scheme can also be used for generating two-mode squeezing lights, if we consider the nano-ellipsoid is optically levitated by dual-tweezers, both A and B.
Driven by the scattering photons, two cavity modes $\hat a_A$ and $\hat a_B$ are excited in the bichromatic cavity.
Based on Equations~\eqref{EQ8},~\eqref{EQ9}, and~\eqref{EQ11}, 
the output cavity mode $\hat a_j^{out}(\omega)$ can be solved by

\begin{equation}
\begin{aligned}
\hat a_j^{out}(\omega )= m_1^j(\omega ){{\hat b}^{in}}(\omega ) + m_2^j(\omega ){{\hat b}^{in\dag }}( - \omega ) + m_3^j(\omega )\hat a_A^{in}(\omega ) 
+ m_4^j(\omega )\hat a_A^{in\dag }( - \omega ) + m_5^j(\omega )\hat a_B^{in}(\omega ) + m_6^j(\omega )\hat a_B^{in\dag }( - \omega )
\label{EQ19}
\end{aligned}
\end{equation}

where
\begin{equation}
\begin{array}{l}
m_1^A = i\sqrt {{\kappa _A\gamma _m}} {g_A}{u_2}{v_2}{w_1}{w_2}/z,\\
\\
m_2^A = i\sqrt {{\kappa _A\gamma _m}} {g_A}{u_1}{v_2}{w_1}{w_2}/z,\\
\\
m_3^A = \kappa _A\left( {\left( {{u_1} - {u_2}} \right)\left( {g_B^2{v_2}\left( {{w_1} - {w_2}} \right) + g_A^2{w_1}{w_2}} \right) + {u_1}{u_2}{v_2}{w_1}{w_2}} \right)/z-1,\\
\\
m_4^A = \kappa _A\left( {g_A^2\left( {{u_1} - {u_2}} \right){w_1}{w_2}} \right)/z,\\
\\
m_5^A = \sqrt {{\kappa _A\kappa _B}} \left( {{g_A}{g_B}\left( {{u_1} - {u_2}} \right){v_2}{w_2}} \right)/z,\\
\\
m_6^A = \sqrt {{\kappa _A\kappa _B}} \left( {{g_A}{g_B}\left( {{u_1} - {u_2}} \right){v_2}{w_1}} \right)/z,
\end{array}
\label{EQ20}
\end{equation}
and
\begin{equation}
\begin{array}{l}
m_1^B = i\sqrt {{\kappa _B\gamma _m}} {g_B}{u_2}{v_1}{v_2}{w_2}/z,\\
\\
m_2^B = i\sqrt {{\kappa _B\gamma _m}} {g_B}{u_1}{v_1}{v_2}{w_2}/z,\\
\\
m_3^B = \sqrt {{\kappa _A\kappa _B}} \left( {{g_A}{g_B}\left( {{u_1} - {u_2}} \right){v_2}{w_2}} \right)/z,\\
\\
m_4^B = \sqrt {{\kappa _A\kappa _B}} \left( {{g_A}{g_B}\left( {{u_1} - {u_2}} \right){v_1}{w_2}} \right)/z,\\
\\
m_5^B = \kappa _B \left( {\left( {{u_1} - {u_2}} \right)\left( {g_B^2{v_1}{v_2} + g_A^2\left( {{v_1} - {v_2}} \right){w_2}} \right) + {u_1}{u_2}{v_1}{v_2}{w_2}} \right)/z-1,\\
\\
m_6^B = \kappa _B \left( {g_B^2\left( {{u_1} - {u_2}} \right){v_1}{v_2}} \right)/z,
\end{array}
\label{EQ21}
\end{equation}
with $z = \left( {{u_1} - {u_2}} \right)\left( {g_B^2{v_1}{v_2}\left( {{w_1} - {w_2}} \right) + g_A^2\left( {{v_1} - {v_2}} \right){w_1}{w_2}} \right) + {u_1}{u_2}{v_1}{v_2}{w_1}{w_2}$.
Furthermore, by introducing two phase-quadrature operators
\begin{equation}
X\left( \omega  \right) = \frac{1}{2}\sum\limits_{j = A,B} {\left[ {a_j^{out}\left( \omega  \right) + a_j^{out\dag }\left( { - \omega } \right)} \right]},
\label{EQ22}
\end{equation}
and
\begin{equation}
Y\left( \omega  \right) = \frac{1}{{2i}}\sum\limits_{j = A,B} {\left[ {a_j^{out}\left( \omega  \right) - a_j^{out\dag }\left( { - \omega } \right)} \right]} ,
\label{EQ23}
\end{equation}
the squeezing spectra $S_{XX}(\omega)$ and $S_{YY}(\omega)$ can be calculated by
$\left\langle {X\left( \omega  \right)X\left( {\omega '} \right)} \right\rangle  = 2\pi {S_{XX}}\left( \omega  \right)\\
\delta \left( {\omega  + \omega '} \right)$
and
$\left\langle {Y\left( \omega  \right)Y\left( {\omega '} \right)} \right\rangle  = 2\pi {S_{YY}}\left( \omega  \right)\delta \left( {\omega  + \omega '} \right)$~\cite{li2015generation}.
The output two-mode cavity field is squeezed if $S_2(\omega) = {S_{XX}}(\omega) + {S_{YY}}(\omega) < 1$.

\begin{figure}
  \centering
  \includegraphics[width=10cm]{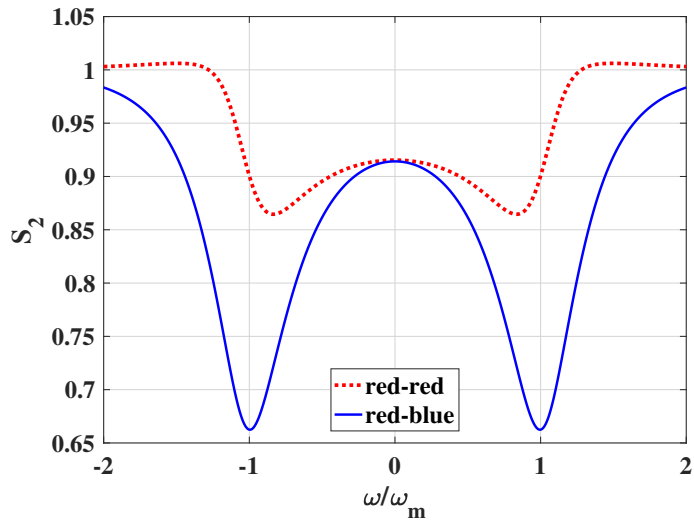}
  \caption{(Color online) Squeezing spectrum of two output modes. The legend red-red represents both two optical tweezers in red sideband of the cavity mode. The legend red-blue denotes the optical tweezers A in red sideband while another tweezers B in blue sideband. The decay rates of the cavity mode are $\kappa_A=0.3\omega_m$ and $\kappa_B=3\omega_m$. Two tweezers are different in wavelength $\lambda_A=780$ nm ($\lambda_B=980$ nm), Other parameters are the same with Figure~\ref{Fig:3}}
  \label{Fig:5}
\end{figure}
\begin{figure}[htbp]
  \centering
\subfigure{ \includegraphics[width=10cm]{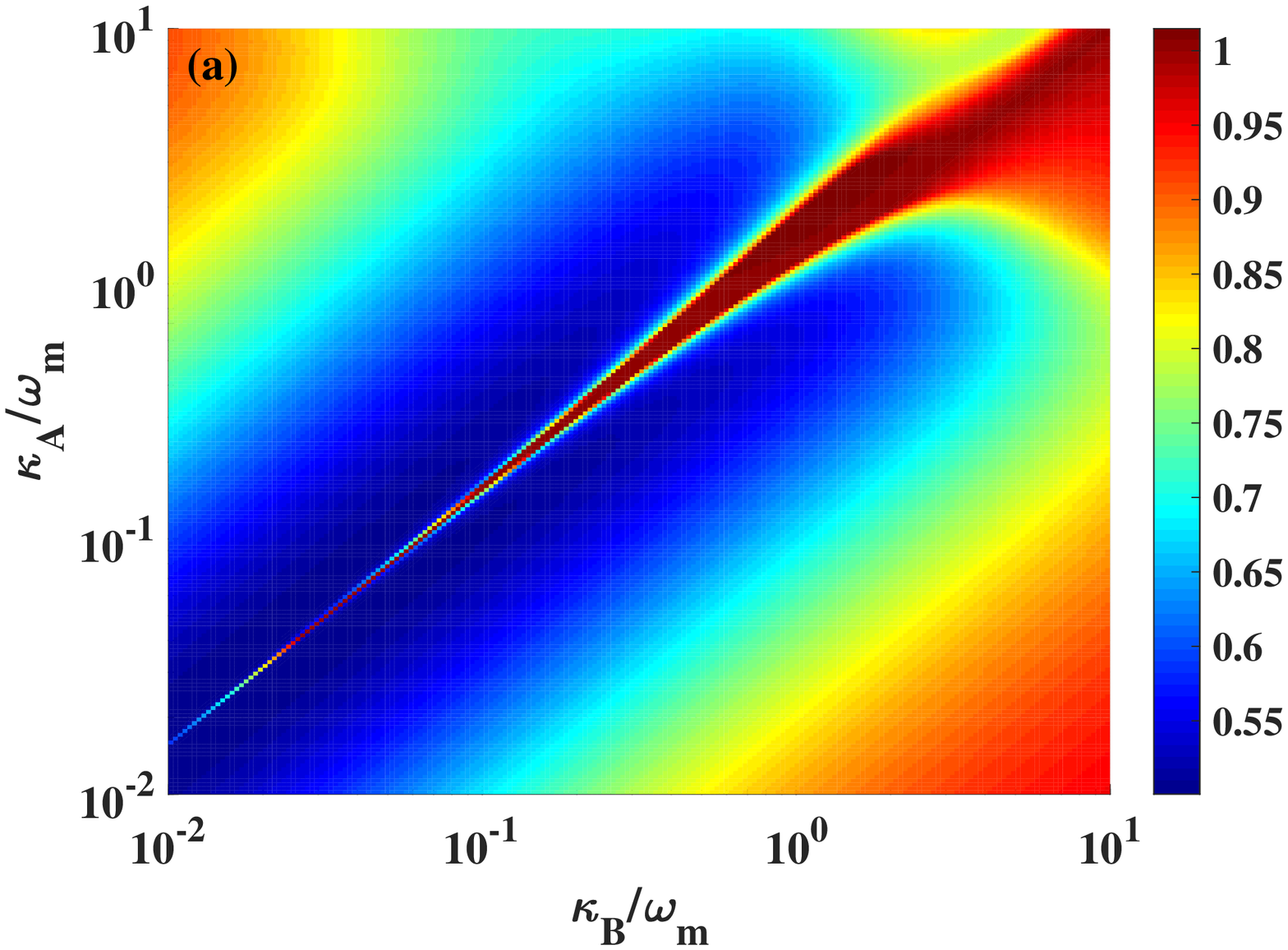}}
\subfigure{  \includegraphics[width=10cm]{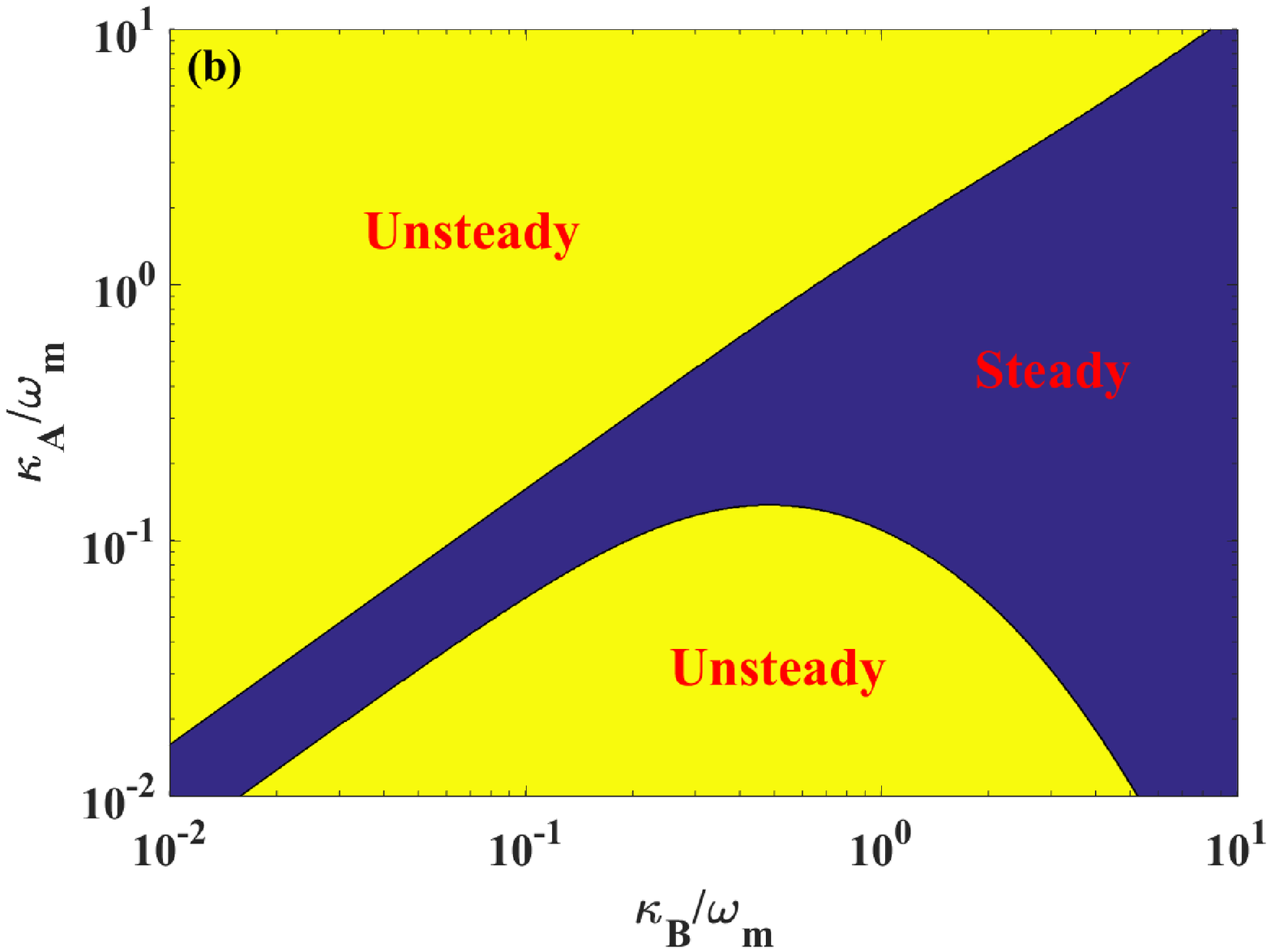}}

\caption{(Color online)  $S_2(\omega)$ (a) and the maximum eigenvalue of Equation~\ref{EQ7} (b) as a function of the cavity decay rate $\kappa_j$. The detunings are set to the red and blue sideband $\Delta_A=-\Delta_B=\omega_m$, respectively. In picture (b), the yellow area means the maximum eigenvalue of Equation~\ref{EQ7} is non-negative. According to the Routh-Hurwitz stability criterion~\cite{vitali2007optomechanical}, the system becomes instable in the long time limit. Conversely, the blue area denotes the system is stable. Other parameters are the same to Figure~\ref{Fig:5}.}
  \label{Fig:6}
  \end{figure}

In order to realize the two-mode squeezing light in steady state, two optical tweezers A and B are detuned to the red sideband (red-red), or the optical tweezers A is detuned to red sideband while another opticla tweezers B is detuned to blue sideband (red-blue).
In the red-blue case, the system will be stable in blue sideband if the cavity photons quickly loss to the vacuum.
So, the cavity decay rates are assumed to $\kappa_B=10\kappa_A=3\omega_m$.
As the torsional mode resonance in blue sideband, the dynamics of the torsional motion will be gradually amplified.
In that case, the enhanced back-action interaction on cavity mode will induce the deeper squeezing.
As shown in  Figure~\ref{Fig:5}, the squeezing of the red-blue case is much larger than that of the red-red case as the mechanical susceptibility of the levitated nano-ellipsoid responds in resonant region.

Next, we will analyze the two-mode squeezing from the view of system stability.
As shown in Figure~\ref{Fig:6}a, none of $S_2(\omega)$ is lower than 0.5 ($3$ dB).
Obviously, there is a narrow and suddenly interrupted area, which splits the blue area into two parts. 
By comparing with Figure~\ref{Fig:6}b, it corresponds to the critical boundary between the stable and instable regions.
The larger $\kappa_B$ and smaller $\kappa_A$ would take the system more stable. 
However, the squeezing will be reduced for the weak back-action.
The maximum $S_2(\omega)$ can approach as the system gets close to the dynamical instability ~\cite{genes2008robust}

\begin{figure}
  \centering
  \includegraphics[width=10cm]{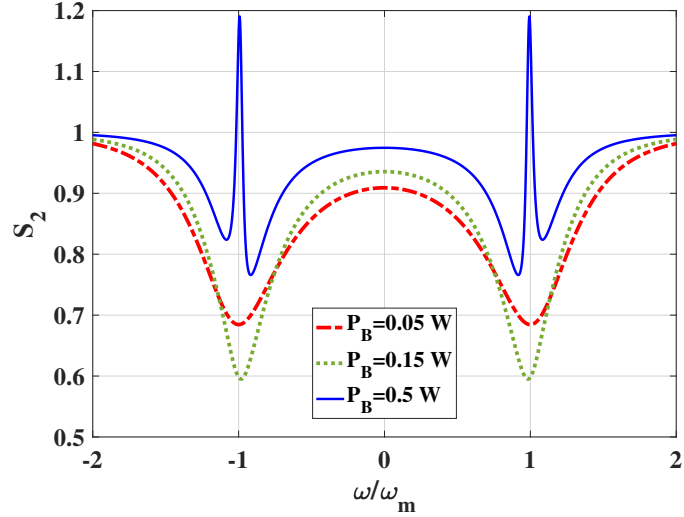}
  \caption{(Color online) $S_2(\omega)$ as a function of $P_B$. The power of optical tweezers A in focus is set to $P_A=0.1$ W.  Tweezers A is detuned to the red sideband while the other tweezers B is set to be the blue sideband. Other parameters are the same as Figure~\ref{Fig:5}}
  \label{Fig:7}
\end{figure}

In Figure~\ref{Fig:7}, we assume that the power of optical tweezers A is fixed at $P_A=0.1$ W.
Noted that the optical tweezers A is set to the red sideband of cavity mode, while optical tweezers B is detuned to the blue sideband.
With the power $P_B$ from 0.05 W to 0.15 W, $S_2(\omega)$ decreases at $\omega \approx \omega_m$ for the enhancement of the back-action interaction.
As the power $P_B$ up to 0.5 W, there is no squeezing on $\omega \approx \omega_m$. 
The amplified torsional motion is overlarge on resonance. 
It results the instability of the system, which shows the destruction of squeezing.
\section {Conclusion} \label{V}

We have proposed the squeezing light source based on an optically levitated nano-ellipsoid, which couples with a cavity.
By the coherent scattering mechanism, the ultra-strong coupling between the cavity field and torsional mode of the nano-ellipsoid could be realized 
under the current experimental parameters.
If the optical tweezers is on the red sideband of the cavity mode,
the single-mode squeezed light source ($17$ dB squeezing)  can be realized even under room temperature.
It is noteworthy that both the strong and broadband squeezing lights can be got under the bad cavity condition. 
In order to achieve the two-mode squeezing lights,
two optical tweezers are applied to trap the nanoparticle at the same time. 
One is on the red sideband, while another is on the blue sideband. The two-mode squeezing of the lights is sensitive to the system stability, which depends on both the cavity decay rates and the power of the optical tweezers.
When the system dynamics closes to the boundary of the stable regime, the two-mode squeezing can be maximized. 
However, two-mode squeezing can not be over $3$ dB limit in our current scheme.  
In future, the $3$ dB limit can be overcame either by feedback~\cite{vinante2013feedback}, or by reservoir-engineering method~\cite{dassonneville2021dissipative}.
\begin{acknowledgements}
This research is supported by National Natural Science Foundation of China under Grant No. 61771278 and Beijing Institute of Technology Research Fund Program for Young Scholars.
\end{acknowledgements}

\appendix
\setcounter{equation}{0}
\setcounter{figure}{0}
\renewcommand{\theequation}{A\arabic{equation}}
\section{Polarization Tensor}
An uniform isotropic non-dispersive ellipsoid is assumed smaller enough than the wave length of the tweezers $\lambda_j$. By this setting, the dipole approximation is performed. The corresponding polarizability of the nano-ellipsoid in a semiaxis $k$ is~\cite{bohren2008absorption}
\begin{equation}
{\alpha _k} = 4\pi abc{\varepsilon _0}\frac{{{\varepsilon _r} - {\varepsilon _0}}}{{3{\varepsilon _0} + 3{L_k}\left( {{\varepsilon _r} - {\varepsilon _0}} \right)}}
\label{A1}
\end{equation}
where $k=\{a,b,c\}$ and ${L_j} = \frac{{abc}}{2}\int_0^\infty  {\frac{{ds}}{{\left( {s + {j^2}} \right){{\left( {s + {a^2}} \right)}^{{1 \mathord{\left/
 {\vphantom {1 2}} \right.
 \kern-\nulldelimiterspace} 2}}}{{\left( {s + {b^2}} \right)}^{{1 \mathord{\left/
 {\vphantom {1 2}} \right.
 \kern-\nulldelimiterspace} 2}}}{{\left( {s + {c^2}} \right)}^{{1 \mathord{\left/
 {\vphantom {1 2}} \right.
 \kern-\nulldelimiterspace} 2}}}}}}
 $.
With $a>b=c$, $L_k$ can be analytically solved by
 \begin{equation}
\begin{array}{l}
{L_a} = \frac{{1 - {e^2}}}{{{e^2}}}\left( { - 1 + \frac{1}{{2e}}\ln \frac{{1 + e}}{{1 - e}}} \right),\\
\\
{L_a} + {L_b} + {L_c} = 1,\\
\\
{L_b} = {L_c}
\end{array}
\label{A2}
 \end{equation}
where $e = \sqrt {1 - {{{b^2}} \mathord{\left/
 {\vphantom {{{b^2}} {{a^2}}}} \right.
 \kern-\nulldelimiterspace} {{a^2}}}} $ denotes the eccentricity.
Then the polarizability tensor for the ellipsoid is given as the matrix product:
\begin{equation}
\hat{ \alpha} = {\hat{R}^{ - 1}}\hat{\alpha}_0 \hat{R},
\label{A3}
 \end{equation}
 where
 $\hat \alpha_0=diag\{L_a,L_b,L_c\}$
and
 $\hat{R} = \left| {\begin{array}{*{20}{c}}
{\cos \theta }&0&{ - \sin \theta }\\
0&1&0\\
{\sin \theta }&0&{\cos \theta }
\end{array}} \right|\left| {\begin{array}{*{20}{c}}
{\cos \phi }&{\sin \phi }&0\\
{ - \sin \phi }&{\cos \phi }&0\\
0&0&1
\end{array}} \right|.
$

\section{Hamiltonian for Coherent Scattering} \label{AP_B}

As the size of ellipsoid is assumed smaller than the wavelength of the optical tweezers, the dipole approximation is appropriate.
The dipole moment is given as
$
\hat P=\hat \alpha(\theta,\phi) \hat E(r,t)
$, 
where $\alpha(\theta,\phi) $ is the polarization tensor characterized by the rotating angles $\theta$ and $\phi$, $\hat E(r,t)$ is the electric field of the incoming beams.
The interaction Hamiltonian between the nano-ellipsoid and electric field can be written as~\cite{gonzalez2019theory,gonzalez2021levitodynamics}
\begin{equation}
{\hat{H}_I} = -\frac{1}{2}\hat{\alpha}(\theta,\phi) {\hat{E}^2(r,t)}
\label{A4}
\end{equation}
where 
\begin{equation}
\hat E(r,t){\rm{ = }}\sum\limits_{j = A,B} {\left( {E_C^j(r) + \varepsilon_T^j(r,t)} \right)}
\label{A5}
\end{equation}
where $E_C^j\left( r \right) = \sqrt {\frac{{\hbar \omega _c^j}}{{2{\varepsilon _0}V_c^j}}} \left[ {{f^j}\left( r \right){{\hat a}_j} + H.c.} \right]$ 
is the electric field in cavity mode and 
$\hat \varepsilon _T^j\left( {r,t} \right) = {{\left[ {E_T^j\left( r \right){e^{i\omega _jt}} + c.c.} \right]} \mathord{\left/
 {\vphantom {{\left[ {E_T^j\left( r \right){e^{i\omega _t^jt}} + c.c.} \right]} 2}} \right.
 \kern-\nulldelimiterspace} 2}$ 
 is the tweezer fields with Gaussian form 
 $E_T^j(r) = {E^j_0}\frac{{w_{t0}^j}}{{w_t^j\left( z \right)}}{e^{ - \frac{{{x^2} + {y^2}}}{{w_t^j\left( z \right)w_t^j\left( z \right)}}}}{e^{i\omega _jt}} \\
 {e^{ik_t^jz}}{e^{i\psi _t^j\left( r \right)}}$.
$\hat a_j$ and $\hat a^{\dag}_j$ are the bosonic annihilation and creator operators of cavity modes.
$\omega_j$ ($\omega^j_c$), $k^j_t$ ($k^j_c$), and $w^j_{t0}$ ($w^j_c$) are the frequency, wave number, and beam waist of the optical tweezers (cavity modes).
$V^j_c=\pi d w^j_c w^j_c/4$ denotes the cavity volume where $d$ is the cavity length.
The mode function $f^j(y)$ depends on the boundary of the cavity, which can be written as $f^j(r)=cos(k^j_c y+\psi_j)$ in $y$ axis.
$E_0^j = \sqrt {{{4P_j} \mathord{\left/
 {\vphantom {{4P_j} {\pi {\varepsilon _0}cw_{t0}^j}}} \right.
 \kern-\nulldelimiterspace} {\pi {\varepsilon _0}cw_{t0}^j}}}$
and $w_t^j\left( z \right) = w_{t0}^j\sqrt {1 + {{{z^2}} \mathord{\left/
 {\vphantom {{{z^2}} {z_R^2}}} \right.
 \kern-\nulldelimiterspace} {z_R^{j2}}}}$
are the amplitude and beam waist of the tweezers. $z^j_R=\pi w^j_{t0} w^j_{t0} /\lambda_j$ and $\psi^j_t(r)$ are the Rayleigh range and Gouy phase of the tweezers.
$P_j^t$, $c$, and $\lambda_j$ are the power, speed, and wavelength of the tweezers.

By inserting Equation~\ref{A5} to Equation~\ref{A4}, the Hamiltonian can be written as $H_I=H_{T-T}+H_{C-C}+H_{T-C}$ where
\begin{equation}
{H_{T - T}} =  - \frac{1}{2} \hat \alpha \left( {\theta ,\phi } \right)\sum\limits_{j = A,B} {\varepsilon _T^j(r,t)\varepsilon _T^j(r,t)},
\label{A6}
\end{equation}
\begin{equation}
{H_{C - C}} =  - \frac{1}{2} \hat \alpha \left( {\theta ,\phi } \right)\sum\limits_{j = A,B} {E_C^j(r)E_C^j(r)},
\label{A7}
\end{equation}
and
\begin{equation}
{H_{T - C}} =  - \hat \alpha \left( {\theta ,\phi } \right)\sum\limits_{j = A,B} {\varepsilon _T^j(r,t)E_C^j(r)}.
\label{A8}
\end{equation}
Then, we assume that the nano-ellipsoid is fixed at the origin $(r=0)$, and only the torsional motion on $\theta$ and $\phi$ is under considered. That is, when $\theta(\phi)\rightarrow0$, one can expand the $H_{T-T}$ to the second order of $\theta(\phi)$. The torsional frequency of the nano-ellipsoid is
\begin{equation}
 {\omega _{\theta \left( \phi  \right)}} = \sqrt {{{\left( {E_0^AE_0^A + E_0^BE_0^B} \right)\left( {{\alpha _a} - {\alpha _b}} \right)} \mathord{\left/
 {\vphantom {{\left( {E_0^AE_0^A + E_0^BE_0^B} \right)\left( {{\alpha _a} - {\alpha _b}} \right)} {2I}}} \right.
 \kern-\nulldelimiterspace} {2I}}}
 \label{A9}
\end{equation}
where $I=M(a^2+b^2)/5$ denotes the inertia of the ellipsoid. $M$ is the mass of the nano-ellipsoid.
The corresponding intrinsic optomechanical coupling and coherent scattering coupling can be derived by expanding the $H_{C-C}$ and $H_{C-T}$ to the first order of $\theta(\phi)$ as $\theta\rightarrow \pi/4, \phi=0 (\phi\rightarrow \pi/4, \theta=0)$, given as
\begin{equation}
 {g^j_{\theta(\phi)} } = \frac{{\left( {{\alpha _a} - {\alpha _b}} \right){\omega^j_c}{\xi^{\theta(\phi)} _0}{{\cos }^2}\left( \varphi_j  \right)}}{{2{\varepsilon _0}{V^j_c}}},
 \label{A10}
\end{equation}
\begin{equation}
 {g^j_{s\theta(s\phi) }} = \left( {{\alpha _a} - {\alpha _b}} \right){E_0}{\xi^{\theta(\phi)} _0}\cos \left( \varphi_j  \right)\sqrt {\frac{{{\omega^j _c}}}{{8\hbar {\varepsilon _0}{V^j_c}}}}
 \label{A11}
\end{equation}
where ${\xi ^{\theta(\phi)} _0} = \sqrt {{\hbar  \mathord{\left/
 {\vphantom {\hbar  {2I{\omega _\phi }}}} \right.
 \kern-\nulldelimiterspace} {2I{\omega _{\theta(\phi)} }}}}$
is the zero-point fluctuation of torsion mode.
It is noted that the torsion motion can be decoupled from the center-of-mass motion by moving the optical tweezers to the anti-node of the cavity mode ($\varphi_j=0)$~\cite{li2021steady}.
Moreover, because ${g^j_{\theta(\phi)} }$ is much smaller than the ${g^j_{s\theta(s\phi) }}$,
the coherent scattering coupling ${g^j_{s\theta(s\phi) }}$ is under considered.
Then the Hamiltonian for the interaction picture can be read by~\cite{gonzalez2019theory}
\begin{equation}
\hat H{\rm{ = }}\hbar \Delta _j\hat a_j^\dag {\hat a_j} + \hbar {\omega _m}{\hat b^\dag }\hat b - \hbar \sum\limits_{j = A,B} {{g_j}\left( {\hat a_j^\dag  + {{\hat a}_j}} \right)\left( {{{\hat b}^\dag } + \hat b} \right)}
\label{A12}
\end{equation}
where $\Delta_j=\omega^j_c-\omega_j$, $\omega_m$ denotes the torsional frequency $\omega_{\theta \left( \phi  \right)}$, and $g_j$ represents the coherent scattering coupling $g^j_{s\theta(s\phi)}$.
In this Hamiltonian, the first two terms denotes the free Hamiltonian for the cavity modes and torsional mode, while the last term describes the coherent scattering interactions between the cavity modes and torsional mode.

\section{Damping for Torsional Motion}

As the nano-ellipsoid optically trapped in a high vacuum environment, the collision of surrounding gas on nano-ellipsoid leads to the change of angular momentum.
The damping rate for the torsional motion is given by~\cite{halbritter1974torque}
\begin{equation}
\begin{aligned}
\gamma_\phi  = \frac{{5{\rho _a}\bar v{a^3}\sqrt {1 - {e^2}} }}{{8\rho \left( {{a^2} + {b^2}} \right)}}\left[ {{\gamma _{ac}}\left( {{f_1} + \left( {1 - {e^2}} \right){f_2}} \right) + 3\left( {1 - {\gamma _{ac}}\frac{{6 - \pi }}{8}} \right){e^4}{f_3}} \right]
\end{aligned}
\label{A13}
\end{equation}
where
\begin{equation}
\begin{array}{l}
{f_1} = \frac{3}{{8{e^2}}}\left[ {\frac{1}{e}\arcsin \left( e \right) - \left( {1 - 2{e^2}} \right)\sqrt {1 - {e^2}} } \right],\\
\\
{f_2} = \frac{3}{{16{e^2}}}\left[ {\left( {1 + 2{e^2}} \right)\sqrt {1 - {e^2}}  - \frac{1}{e}\arcsin \left( e \right)\left( {1 - 4{e^2}} \right)} \right],\\
\\
{f_3} = \frac{1}{{4{e^4}}}\left[ {\left( {3 - 2{e^2}} \right)\sqrt {1 - {e^2}}  + \frac{1}{e}\arcsin \left( e \right)\left( {4{e^2} - 3} \right)} \right].
\end{array}
\label{A14}
\end{equation}
${\rho _a} = {{{m_a}{p}} \mathord{\left/
 {\vphantom {{{m_a}{p_a}} {{k_B}}}} \right.
 \kern-\nulldelimiterspace} {{k_B}}}{T_a}$
is the mass density, where $m_a$, $p$, and $T_a$ are atom mass, pressure, and temperature of the residual gas.
$\bar v = \sqrt {{{8{k_B}{T_a}} \mathord{\left/
 {\vphantom {{8{k_B}{T_a}} {\pi {m_a}}}} \right.
 \kern-\nulldelimiterspace} {\pi {m_a}}}} $ is the mean thermal velocity and $\rho$ is the density of the nano-ellipsoid.
$ {\gamma _{ac}}$ denotes the accommodation efficient which charges the diffuse and specular reflection of ellipsoidal surface.

\bibliography{main.bib}
\end{document}